\documentstyle[prl,twocolumn,aps,epsf,epsfig]{revtex}
\begin{document}
\draft
\twocolumn[\hsize\textwidth\columnwidth\hsize\csname@twocolumnfalse%
\endcsname

\title{Skyrmion Strings and Anomalous Hall Effect in Double Exchange Systems.}

\author{M.J. Calder\'on and L. Brey}
\address{Instituto de Ciencia de Materiales de Madrid, 
CSIC,
28049 Cantoblanco, Madrid, Spain.}

\date{\today}

\maketitle

\begin{abstract}
\baselineskip=2.5ex
We perform Monte Carlo simulations to obtain quantitative results for the 
anomalous Hall resistance, $R_A$, observed in colossal magnetoresistance manganites. 
$R_A$ arises from the interaction between the spin 
magnetization and topological defects via spin-orbit coupling.
We study these defects and how they are affected by the spin-orbit 
coupling within the framework of the double exchange model. 
The obtained anomalous Hall resistance is, in sign, order of magnitude and shape, 
in agreement with experimental data.     

\end{abstract}
]

Doped perovskite manganites have attracted much attention lately, since 
they show the so called colossal magnetoresistance\cite{review}.
These materials undergo a ferromagnetic-paramagnetic transition accompanied
by a metal-insulator transition.
The Double-Exchange (DE) mechanism\cite{detheory} plays
a major role to explain this magnetic transition. In the DE picture,
the carriers moving through the lattice are strongly ferromagnetically coupled
to the $Mn$ core spins and this produces a modulation of the hopping amplitude
between neighboring $Mn$ ions.

The Hall resistivity $\rho _H$ in ferromagnets has two contributions\cite{hurd},
one proportional to the magnetic field ${\bf H}$, and the other 
to the spin magnetization ${\bf M}$:
$\rho _H$=$R_O H+ R_A M$.
$R_O $ and $ R_A $ denote the ordinary (O) and the anomalous (A) 
Hall resistances (HR).
The existence of $R_A$ requires a coupling of orbital motion of electrons
to ${\bf M}$, and the AHE  is usually explained in terms of skew scattering
due to spin-orbit (s-o) interaction\cite{nozieres}.

Several groups\cite{matl,chun,ziese,chun1} have measured $\rho _H$
of the doped $Mn$ oxide $La_{0.7}Ca_{0.3}MnO_3$ for different temperatures ($T$).
These experiments found that
$R_O$ and $R_A$ have opposite sign, that $R_A$ is much bigger than $R_O$,
that $R_A$ peaks at a $T$ above $T_c$, and decreases 
slowly at higher  $T$.
These effects can not be explained with the conventional skew scattering theory\cite{ybk}.
Recently, it was proposed\cite{ye,lyanda}that in $Mn$ oxides $R_A$ arises from the
interaction of ${\bf M}$ with non trivial spin textures (topological charges)
via s-o coupling.
The number of topological charges in the three-dimensional (3D) Heisenberg 
model increases exponentially with $T$\cite{Hdefects} and  
Ye {\it et al.}\cite{ye} extrapolated these results to
DE materials obtaining a rapid increase of $R_A$ at low $T$.
For $T > T_c$, although in 3D there is not a theory of topological defects, 
Ye {\it et al.}
were able to estimate the overall shape of $R_A$.

In this work we perform Monte Carlo simulations in order to obtain a
quantitative form of $R_A$ in DE systems.
We compute the number of topological defects as a function of $T$ and,
by introducing s-o interaction, we couple the defects orientation
with ${\bf M}$. In this way we obtain an AHR which has the same sign, 
shape and order of magnitude as found experimentally\cite{matl}.

{\it Hamiltonian.}
The electronic and magnetic properties of the $Mn$ oxides
are described by the 
DE Hamiltonian, 
\begin{eqnarray}
\hat H    =  &  - & t \sum _{i \ne j, \sigma} 
e^{ i a { e \over {\hbar c}} A _{i {\hat \delta}}} \,
d ^+ _{i,\sigma}  d  _{j,\sigma}
\nonumber \\ &  - & 
J_H \sum _{i,\sigma,\sigma '} d ^+ _{i,\sigma} {\mbox {\boldmath $\sigma$}} _{\sigma,\sigma'}
d_{i,\sigma'} {\bf S}_i + {\hat H}_{so}  + {\hat H}_z \, \, \, ,
\end{eqnarray}
here $ d ^+ _{i,\sigma}$ creates an electron at site $i$ and with spin $\sigma$,
$t$ is the hopping amplitude between nearest-neighbor sites,  
$J_H$ is the Hund's rule 
coupling energy, 
${\bf S}_i$ is the core spin at site $i$ and ${\hat H} _{so}$ is the spin-orbit (s-o)
interaction.
In the tight binding approximation, the effect of  ${\bf H}$ is to modify
the hopping matrix element by introducing the phase 
$ a { e \over {\hbar c}} A _{i {\hat \delta}}$, where
$a$ is the lattice parameter, $j=i+ {\hat \delta}$ and ${\bf A}$ is the vector potential
corresponding to ${\bf H}$.
The last term in the Hamiltonian is the Zeeman coupling
${\hat H} _z =  g \mu _b {\bf H}  \sum _ i {\bf S}  _ i $.
We assume that the $Mn$ ions form a perfect cubic lattice. 

In the limit of infinite $J_H$, the electron spin at site $i$ should be parallel
to  ${\bf S}_i$, and ${\hat H}$ becomes\cite{detheory}
\begin{equation}
\hat H \! = \! -  t   \sum _{i \ne j } \! \! 
\cos {{{\theta _{i,j}} \over 2}}
e ^ {i \left ( \phi (i,j)/2 +  a { e \over {\hbar c}} A _{i {\hat \delta}} \right )}  
c^+ _i c_j \! 
\! + \!  {\hat H} _{so} \! \! + \! \! {\hat H} _z.
\end{equation}
Now $c _i ^+ $ creates an electron at site $i$ with spin parallel to
${\bf S } _i$,   ${\bf m} _i$=$ {\bf S } _i / S$ and 
$\cos {\theta _{i,j}}$=${\bf m}_i \! \cdot \! {\bf m}_j $.
The first term in {\mbox Eq.(2)} describes the motion of electrons
in a background of core spins. 
The electron hopping is affected by the {\mbox nearest} neighbor spin overlap,
$\cos (\theta _{i,j} / 2)$,
being the kinetic energy minimum when all core spins are parallel. 
This is the DE mechanism for the existence of a
ferromagnetic metallic  ground state in $Mn$ doped oxides.
When the orientation of the electron spin is moved around a close loop the
quantum system picks up a Berry's phase\cite{berry}
proportional to the solid angle enclosed by the tip of the spin
on the unit sphere. $\phi (i,j)$ is  the
Berry's phase defined mathematically as the solid angle subtended 
by the unit vectors ${\bf m} _i$, ${\bf m} _j$ and $\hat z$ on the unit sphere.

The phase $\phi (i,j)$ in the Hamiltonian affects the motion of electrons 
in the same way as does 
the phase arising from a physical magnetic field.
$\phi (i,j)$  is related with internal gauge fields generated by non trivial
spin textures, which appears in the system when increasing T.
In absence of s-o coupling the phases $\phi (i,j)$ are random
and the net internal gauge field is zero.
In the presence of s-o coupling
there is a privilege orientation for the spin textures and a non
zero average internal gauge field appear.

In the following we study, as a function of T, the appearance
of topological defects in the DE model.
Once the spin textures are characterized, we study the effect that 
the s-o coupling
has on the defects and we analyze their
contribution to the Hall effect.

{\it Topological defects in the DE model.}
The temperature induces fluctuations in the ferromagnetic state
of the core spins and at  $T > T_c$
the system becomes  paramagnetic.
Typically $T_c \sim  300K$ and this $T$ is much smaller that the 
electron Fermi energy, $t \sim  0.1eV$,  therefore we consider
that the conduction electrons temperature is zero.

In first order in the  electron wave function the temperature dependence of the 
magnetic properties is described by the classical action\cite{mj1},
\begin{equation}
S = \beta \, t \, { \langle c^+ _i c_ j \rangle} _0 \sum _{i \ne j} 
\cos {{{\theta _{i,j}} \over 2}} \, 
e ^ {i \phi (i,j)/2} 
+ {\hat H}_z
\, \, \, ,
\end{equation}
where $\langle \,\,\, \rangle _0$ means expectation value at $T=0$.

The system described by Eq.(3) is expected to behave with $T$ similarly to the 
Heisenberg model.
In particular, at low $T$,  
we expect the occurrence of  
point
singularities as those occurring  in the 3D Heisenberg model\cite{Hdefects}.
These points singularities are called topologically stable because no local
fluctuations in a uniform system can produce them.
The  defects  are  classified\cite{Hdefects}
by the topological charge, $Q$, which represents the number
of times and the sense in which spins on a closed surface
surrounding the defect cover the surface of a unit sphere in spin space.
An example of topological defects with 
$Q = +1 (-1) $ occurs at a position from where
all the spins point radially outward (inward).

In order to locate the topological charges in the lattice model we follow the prescription
of Berg and L\"{u}sher\cite{berg}.
For each unit cube of the lattice we divide each of its six faces into two triangles.
The three unit-normalized spins at the corners of a triangle $l$
define a signed area $A_l$ on the unit sphere. The topological charge 
enclosed by the unit cube is given by
\begin{equation}
Q = {1 \over {4 \pi}} \sum _{l=1} ^{12} A _l \, \, \, \, .
\end{equation}
This definition of $Q$ ensures that, 
when using periodic boundary conditions,
the total topological charge in the system  
is  zero. 
$Q$ in each cube is an integer number,
and the magnitude of nonzero charges is almost always equal
to unity. Only at rather high temperatures a  few defects with $|Q|\ge 2$ are found.

$Q$ can be interpreted 
as the number of magnetic monopoles enclosed by the unit cube.
The quantities $ A _l  \phi _ 0 / {4 \pi}$ 
represent the magnetic flux piercing the
triangle $l$. Here $\phi  _0 = hc /e$ is 
the magnetic flux quantum. 
Following this analogy we  assign at each point of the lattice, $i$,  a three
dimensional internal magnetic field ${\bf b}_i$.

The phase $\phi (i,j)$ can  be written in the form,
\begin{equation}
{\phi (i,j) \over{2}}= a  { e \over {\hbar c }}   a_{i{\hat \delta}} \, \, ,
\end{equation}
where ${\bf b}$=${\bf \nabla} \times {\bf a}$.
Here it is clear that the  ${\bf b}$ 
associated with a  spin texture affects  
the motion of an electron just as does an external
magnetic field.

In a system with a uniform magnetization at the surface, the interaction
between a positive and a negative defect is finite and in the continuous
Heisenberg model increases linearly with separation.\cite{ostlund}
Defects with opposite charges are closely bound in pairs at 
low $T$. These pairs of defects are Skyrmion\cite{rajaraman,comment}
strings (dipoles) 
which begin at a monopole ($Q$=1) and end at an antimonopole ($Q$=-1).
The Skyrmions are characterized by a dipole ${\bf P}$ joining $Q$=-1, and
$Q$=+1, 
\begin{equation}
{\bf P} =  {{a ^2}  \over {\phi _0 } } \sum _i {\bf b} _i \, \, \, .
\end{equation}

\begin{figure}
\epsfig{file=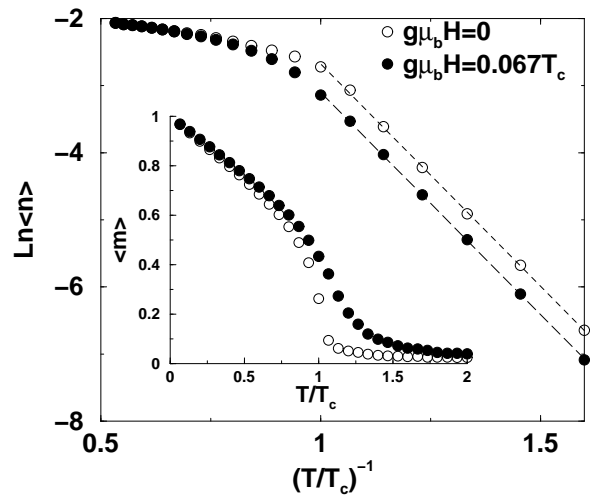,width=3in}
\caption{Plot of the logarithm of the average defect pair density as a 
function of 
the inverse temperature for a system of size $16 \! \times \!16 \! \times \! 16$. 
The dashed lines are linear fits to the data below $T_c$. Their slope is the core 
energy $E_c=7.05T_c$.}
\label{logn}
\end{figure}

By performing Monte Carlo simulations on the classical variables ${\bf m}_i$
\cite{mj1}, we have studied the 
dependence on $T$ of the number of defects 
in the system. With the definition of $Q$, the number of positive and negative
defects is the same 
and the important quantity is the average defect pair
density $\langle  n  \rangle$.
In Fig. \ref{logn} we plot $\langle  n  \rangle$ as a function of $T$ for a DE system of size
16$\times$16$\times$16. We have checked that our results are free of
finite size 
effects. 
In the same figure we plot  $m$(T).
In the DE system   $T_c \sim 1.06\, t  \langle  c^+ _ i c _j  \rangle_0$.
The results correspond to  $g \mu_b H$=0 and  $g \mu_b H$=0.067$T_c$ ($H\sim$ 10 Tesla). 

We have  studied  correlation between 
defects. For $T < T_c$ 
the $Q$=+1 and $Q$=-1 defects are strongly coupled forming Skyrmions.
The Skyrmions are very dilute, almost independent, and their density 
can be fitted to
\begin{equation}
\langle n \rangle = \alpha e ^{ - \beta E_c} \, \, \, .
\end{equation}
Here $E_c$ and $\alpha$ are, respectively, the core energy and the 
entropy of the Skyrmions.
Numerically we have obtained $E_c = 7.05 T_c$. This value is slightly smaller than
the value obtained for the Heisenberg model\cite{Hdefects} $E_c ^{Heis}$ = 8.7 $T_c$.
At $T > T_c$ the number of defects increases sharply and it becomes very difficult
to pair up defects with opposite charges in an unambiguous way.
The core energy practically does not depend on  $H$, this result
is in agreement with the obtained for pure two-dimensional Skyrmions\cite{fertig}.
The entropy $\alpha$ is related with the degeneracy of the Skyrmions
in the orientation of ${\bf P}$ and with
thermally activated twist and dilatation soft modes.
We have obtained that for $H$=0, $\alpha$=51. Note that $\alpha^{Heis} \sim 320$
\cite{Hdefects}. For finite ${\bf H} $ 
the six (1,0,0)
directions keep being degenerated, but some of the twist soft modes
becomes more gaped and the value of the entropy is reduced. 

By performing 
calculations  in a 
system constrained to get a unique  Skyrmion  with a given  dipole ${\bf P}$,
we have calculated the energy of isolated Skyrmions. 
In this way we have checked that in the absence of s-o coupling
the energy of the Skyrmion only depend on the absolute value of ${\bf P}$
and not on its orientation. We have obtained that the core energy of a Skyrmion
with ${\bf P}$=(1,0,0) is $E_c  ^{ (1,0,0)}\!$ =7.05$T_c$, 
with ${\bf P}$=(1,1,0) is $E_c  ^{ (1,1,0)}\!$ =8.65 $T_c$
and
for 
{\bf P}=(1,1,1) is $E_c  ^{ (1,1,1)}\!$ =10.11 $T_c$. The dependence of the energy
on $P$ confirm the strong confinement  energy of the topological
defect\cite{ostlund}. The increase of the Skyrmion energy with $P$, implies 
that for $T < T_c$ the only relevant pairs are those with $P$=1.

In the case of zero s-o coupling, the Skyrmion core energy does not
depend on the direction of 
${\bf P}$ and the average internal 
gauge magnetic field is zero, $\langle{\bf b}\rangle$=0. 
The s-o coupling privileges  an orientation of the 
Skyrmions and it results in a finite  value of $\langle{\bf b}\rangle$. 

{\it Spin-orbit interaction.} The s-o coupling  has two contributions,
the interaction of the spin of the carriers with the ion
electric field, and the magnetic interaction between the carriers with the
core spins ${\bf S}_i$. In the $J_H \rightarrow \infty$ limit, both contributions have the same 
form\cite{ybk,ye} and, 
\begin{equation}
{\hat H}_{so}=\lambda _ {so} { x \over 2} { {a^2} \over {\phi_0}} S
\sum _i {\bf m}_i \cdot {\bf b} _i \, \, \, ,
\end{equation}
where $x$ is the carriers concentration.
For $\lambda _{so} \ne 0$ and $m \ne 0$,  the Skyrmions
are preferentially	 
oriented.  If {\mbox{ ${\bf H} \parallel {\hat z}$}}, ${\bf m} \parallel {\hat z}$
and a net $z$ component
of $\langle{\bf b}\rangle$ results.

\begin{figure}
\epsfig{file=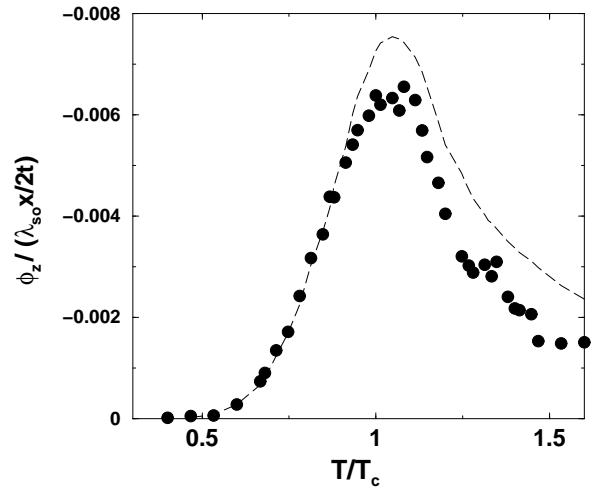,width=3in}
\caption{Plot of $\phi _z$ as a function of $T$ for a system of size $12 \! \times
 \!  12 \!  \times  \! 12$ and $g \mu_b H=0.067T_c$. 
The interaction of the spin magnetization and 
topological defects via spin-orbit coupling has been included. The dashed line
corresponds to an independent Skyrmions picture (see Eq.\ref{indep}).}
\label{bz}
\end{figure}

We have included the s-o interaction in the Monte Carlo simulation and
we have obtained that $\langle b_z \rangle$ is linear with $\lambda _{so}$ in all
range of $T$. 
In Fig. \ref{bz} we plot 
for a system of size 
$12 \! \times \!  12 \!  \times  \! 12$, 
$\phi _z$=$\langle b_z \rangle a ^2 /\phi _0$ as a function of $T$. 
For $T < T_c$ our results are
free of finite
size effects. For $T > T_c$ the increase of the system size does not 
change the overall shape.

It is interesting to analyze the results for $\phi _z$  as a function
of the density of defects $\langle n \rangle$. The main {\mbox effect} of the s-o coupling is to
privilege the appearance of Skyrmions polarized parallel to 
${\bf m}$, $\langle n_- \rangle$, 
with respect to Skyrmions with ${\bf P}$ antiparallel to 
${\bf m}$, $\langle n _+ \rangle$.
This asymmetry produce a $z$-component of the internal gauge field,
$\langle b_z  \rangle$=$\phi_0 / a^2 (\langle  n_-  \rangle\! - \! \langle  n_+  \rangle)$.
Assuming that the Skyrmions are independent, 
\begin{equation}
\langle n_- \rangle - \langle n_+ \rangle= -{1 \over 3} \langle n \rangle \sinh ( \varepsilon _0 /T) \, \, \, ,
\label{def}
\end{equation}
where 
\begin{equation} 
\varepsilon _0 ={{\lambda _{so} x } \over 2} {\tilde m } (T) \, \,  \, ,
\end{equation}
is the s-o energy interaction
of a Skyrmion with ${\bf P}= (0,0,1)$ and
$ {\tilde m } (T)$ is the spin polarization  inside the
Skyrmion. Finally, within the independent Skyrmions picture,
after linearizing Eq.\ref{def},
\begin{equation}
\phi _z ^{ind} = - {{\langle n \rangle } \over  6} 
{{\lambda _{so} } \over  T} x \, {\tilde m} (T) \, \, \, .
\label{indep}
\end{equation} 
${\tilde m} (T)$ is expected to be a function of 
$m(T)$. By comparing the expression for $\phi _z ^{ind}$ with the
Monte Carlo results, Fig. \ref{bz}, we obtain that for $T \leq T_c$, 
${\tilde m } (T)$=$m(T)/5$. That {\mbox means}
that the spin polarization inside the Skyrmions is  five times smaller that
the average spin polarization.  It is interesting to note how the independent
Skyrmions picture describes not just the low temperature regimen, but also
the trends of $\langle b_z  \rangle$ at $T \geq T_c$.

The sign of $\phi _z $ depends on the sign of $\lambda _{so}$. Physically 
we expect that the motion of the electrons leads to an
internal gauge field which acts to cancel the applied field, and
therefore a negative sign for  $\langle b_z  \rangle$ is expected.

\begin{figure}
\epsfig{file=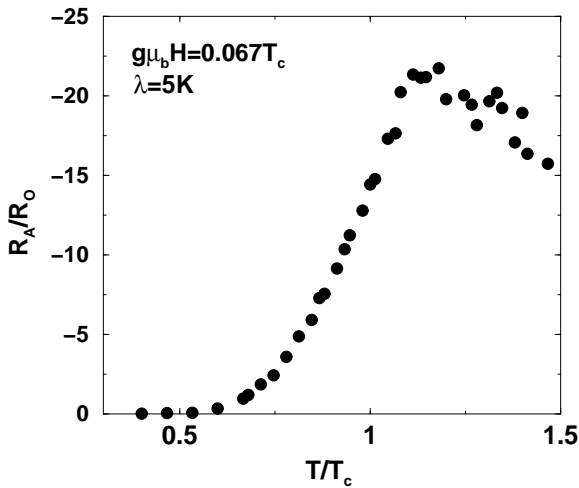,width=3in}
\caption{Plot of $R_A/R_O$ as a function of temperature. This is our main result.
We recover the sign, order of magnitude and shape of the measured anomalous Hall
resistance.}
\label{RaRo}
\end{figure}

{\it Anomalous Hall effect.} The Hall resistivity can be written
as $\rho _H = (H+\langle b \rangle)/nec$. Comparing this expression with 
the definition of $R_A$ we obtain
\begin{equation}
{{ R_A } \over {R_O}} = - { { a^3} \over {g \mu _b}}  { { \langle b_z \rangle} \over {m}}  .
\end{equation}
By using the appropriate parameters, we plot in Fig.  \ref{RaRo} $R_A/R_0$ for
$La_{0.7}Ca_{0.3} Mn O_3$. We take $T_c \sim  270K$\cite{matl}, we use 
$\lambda_{so}$=5$K$\cite{ye} and $g \mu_b H=0.067T_c$.

Fig.3 is the main result of our calculation. We obtain  an  AHR which 
i) is negative, ii) increases exponentially and becomes evident at $T \sim T_c/2$, 
iii)at temperatures close to $T_c$, is around 20 times bigger than the
ordinary Hall resistance, and iv) has a maximum at temperatures {\mbox slightly}
higher ($\sim 30K$) than $T_c$. The subsequent decrease is due to thermal 
fluctuations that destroy the directional order of the Skyrmions. If we take 
into account the fact that the conductance at $T > T_c$ is mainly due to 
polaron hopping, an extra factor of $1/T^2$ is expected\cite{ye} leading to a steeper decrease.
Our results are in good agreement with the data obtained
experimentally\cite{matl,chun}.
We have checked that these results do not depend significantly on the 
magnetic field applied.

In addition, we have found, by diagonalizing the electron Hamiltonian, that there
is not significant electronic charge associated to the Skyrmions\cite{mj_un}. 
This is in contrast to the quantum Hall ferromagnetic systems where 
the topological and electrical charge are equivalent\cite{fertig}.
The Skyrmions we have studied appear when increasing $T$ but we have found that 
they can also appear at $T=0$ when spin defects (in particular, antiferromagnetic islands) 
are present in the system\cite{mj_un}.

In closing, we have computed the anomalous Hall resistance in colossal
magnetoresistance manganites. We have obtained, as a function of temperature,
the average value of Skyrmions for the double-exchange model.
By introducing a spin-orbit interaction,  we obtain a coupling between the 
Skyrmions orientation and the magnetization which results in the 
appearance of an anomalous Hall resistance. The sign, order of magnitude
and shape of the obtained anomalous Hall resistance are in agreement
with the experimental information.

This work was supported by the Cicyt of Spain under Contract No. PB96-0085
and by  the
CAM under Contract No. 07N/0027/1999.


\begin{references}
\bibitem{comment}For simplicity, we will refer to Skyrmion strings simply as Skyrmions.
\end{references}
\end{document}